\documentclass{sig-alternate}
\toappear{~}
\usepackage{graphicx}
\usepackage{balance}  %

\usepackage[utf8]{inputenc}
\usepackage{array}
\usepackage{multirow}
\usepackage{amssymb}
\usepackage{booktabs}
\usepackage{url}
\usepackage[caption=false]{subfig}
\usepackage{xspace}
\usepackage{xcolor}

\usepackage{lmodern}
\usepackage[T1]{fontenc}
\usepackage{microtype}

\newcommand{\eg}{e.\,g.\xspace}
\newcommand{\ie}{i.\,e.\xspace}
\newcommand{\cf}{c.\,f.\xspace}
\newcommand{\dib}[1]{}

\newcommand{\arccost}{\ensuremath{c}}
\newcommand{\queryweight}{\ensuremath{w}}

\newcommand{\hide}[1]{}

\begin{document}

\title{Fast Exact Shortest Path and Distance Queries on Road Networks with Parametrized Costs}

\numberofauthors{3}
\author{
\alignauthor
Julian Dibbelt\\
       \affaddr{Karlsruhe Institute of Technology}\\
       \affaddr{Am Fasanengarten 5}\\
       \affaddr{76131 Karlsruhe, Germany}\\
       \email{dibbelt@kit.edu}
\alignauthor
Ben Strasser\\
       \affaddr{Karlsruhe Institute of Technology}\\
       \affaddr{Am Fasanengarten 5}\\
       \affaddr{76131 Karlsruhe, Germany}\\
       \email{strasser@kit.edu}
\alignauthor 
Dorothea Wagner\\
       \affaddr{Karlsruhe Institute of Technology}\\
       \affaddr{Am Fasanengarten 5}\\
       \affaddr{76131 Karlsruhe, Germany}\\
       \email{dorothea.wagner@kit.edu}
}

\date{30 July 1999}

\maketitle

\begin{abstract}
We study a scenario for route planning in road networks, where the objective to be optimized may change between every shortest path query.
Since this invalidates many of the known speedup techniques for road networks that are based on preprocessing of shortest path structures, we investigate optimizations exploiting solely the topological structure of networks.
We experimentally evaluate our technique on a large set of real-world road networks of various data sources. With lightweight preprocessing our technique answers long-distance queries across continental networks significantly faster than previous approaches towards the same problem formulation.
\end{abstract}

\begin{figure}
\begin{centering}
\subfloat[\label{fig:OSM-Input}OSM Input]{\begin{centering}
\includegraphics[scale=0.137]{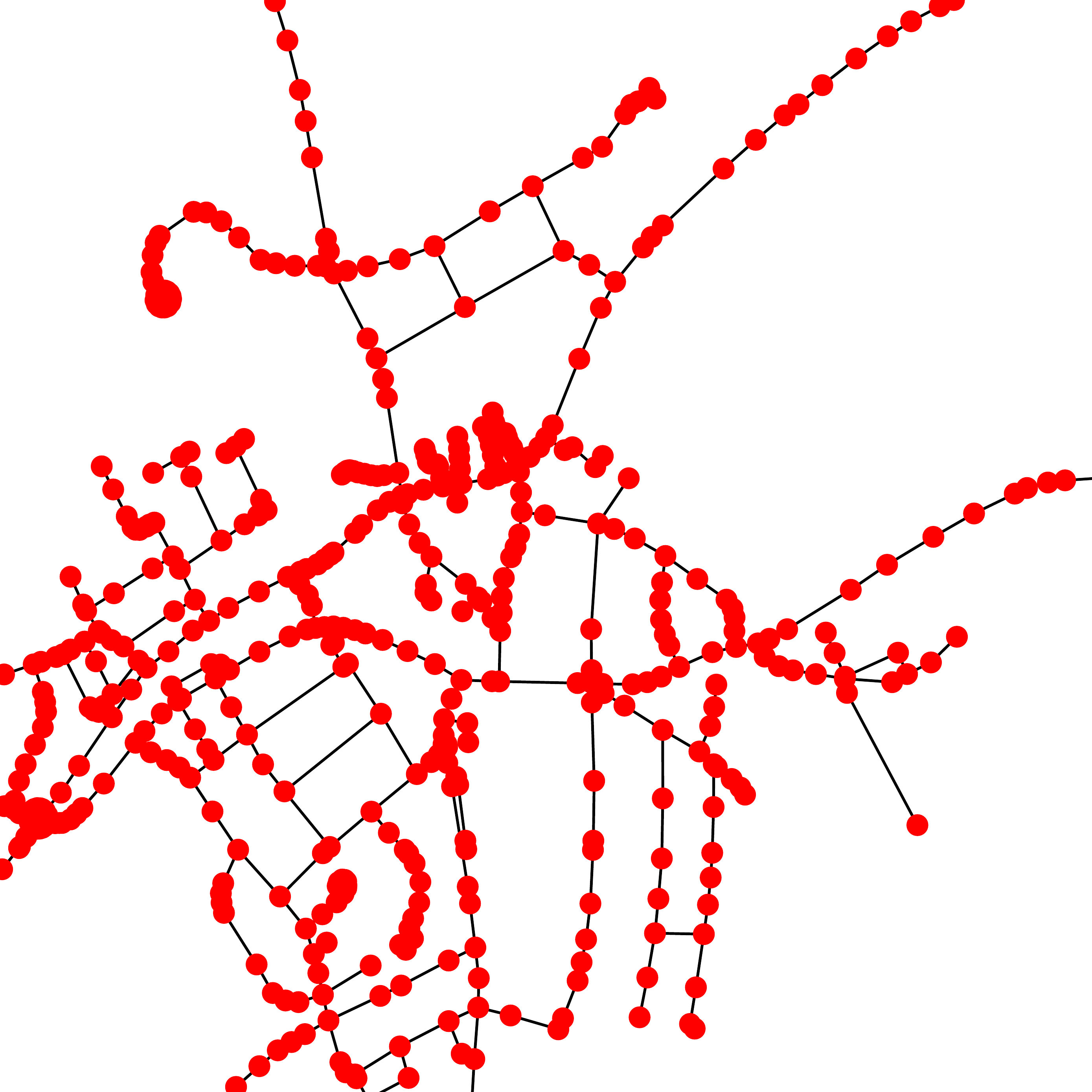}
\par\end{centering}

}\subfloat[\label{fig:DIMACS-Input}DIMACS Input]{\begin{centering}
\includegraphics[scale=0.137]{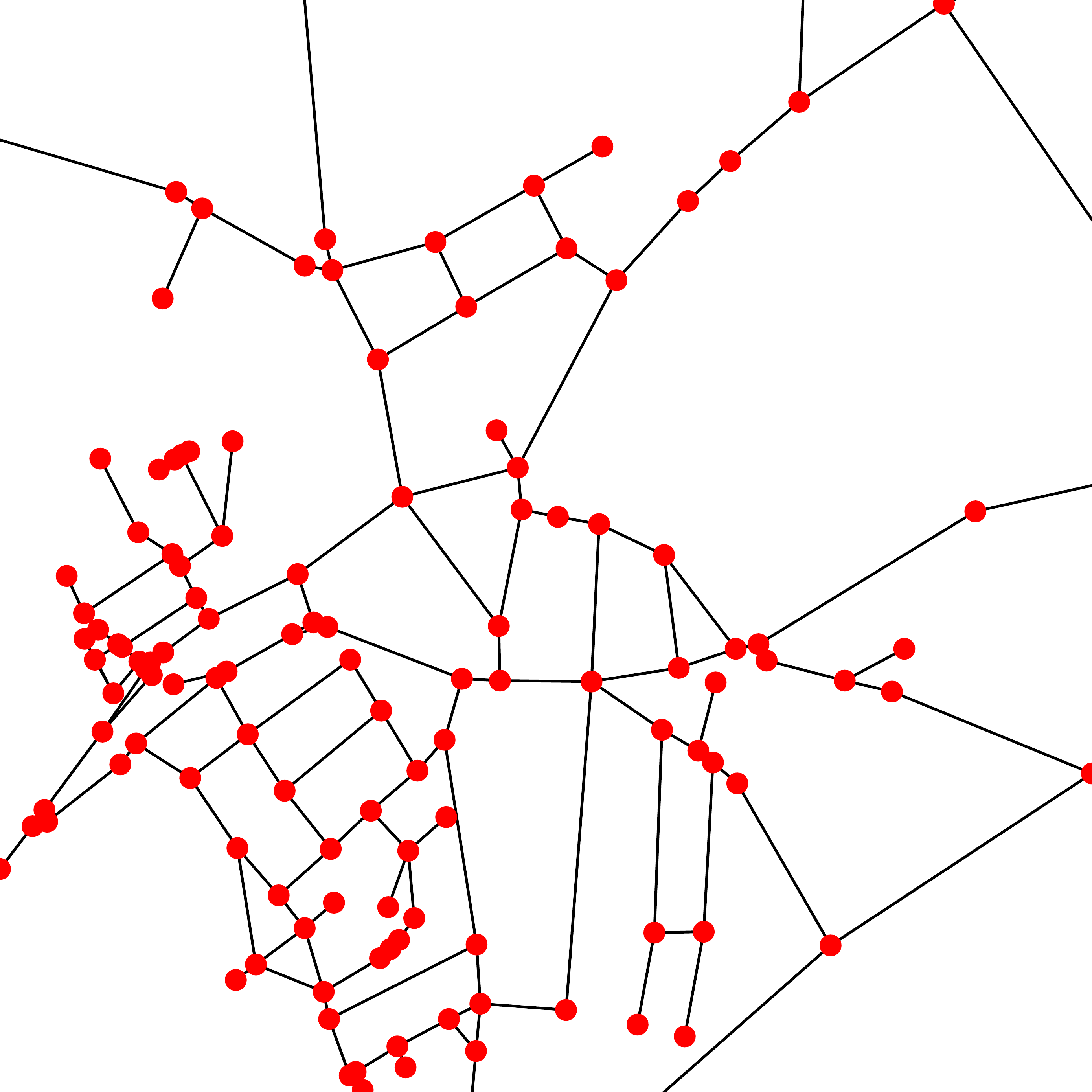}
\par\end{centering}

}
\par\end{centering}

\begin{centering}
\subfloat[OSM Biconn.\ Comp.]{\begin{centering}
\includegraphics[scale=0.137]{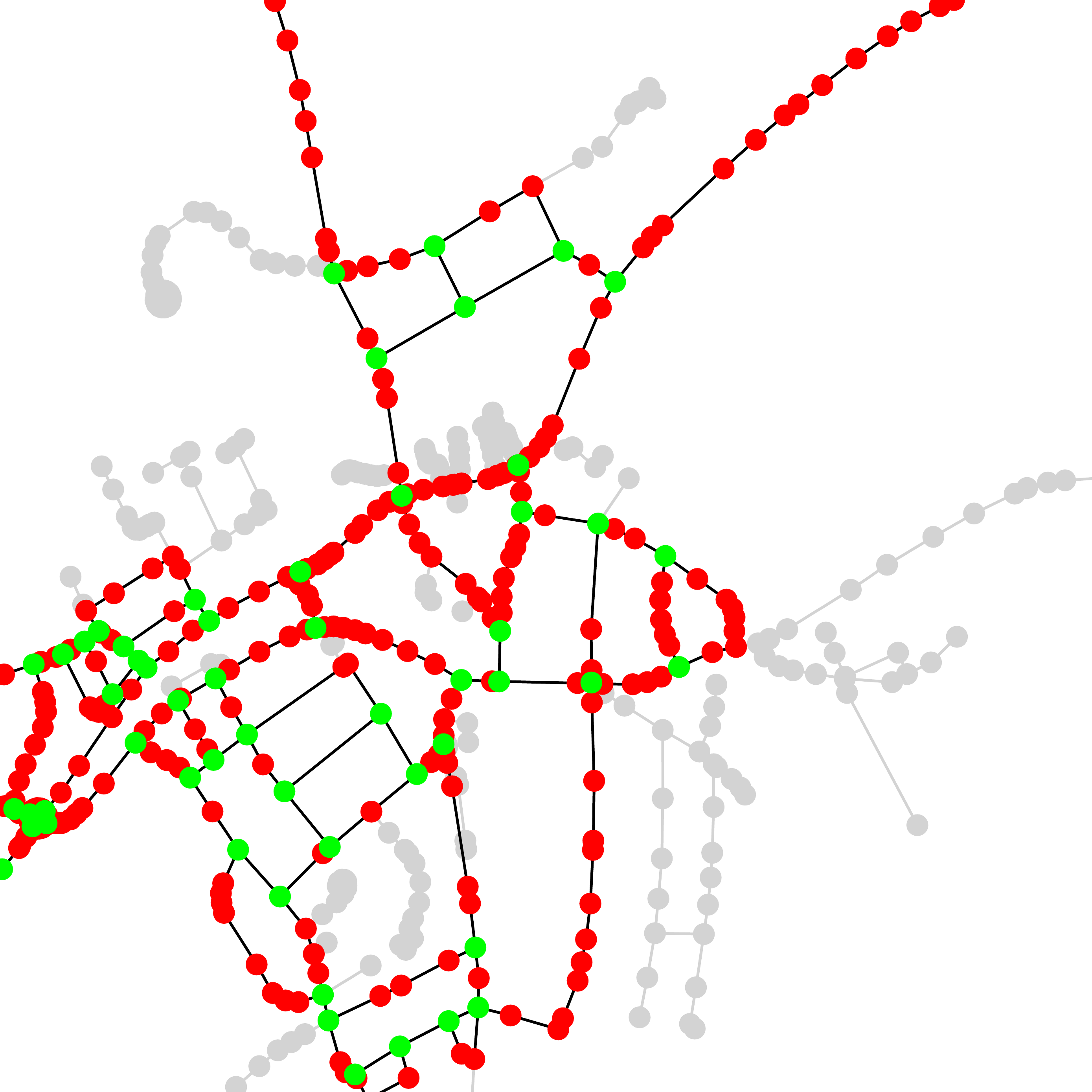}
\par\end{centering}

}\subfloat[DIMACS Biconn.\ Comp.]{\begin{centering}
\includegraphics[scale=0.137]{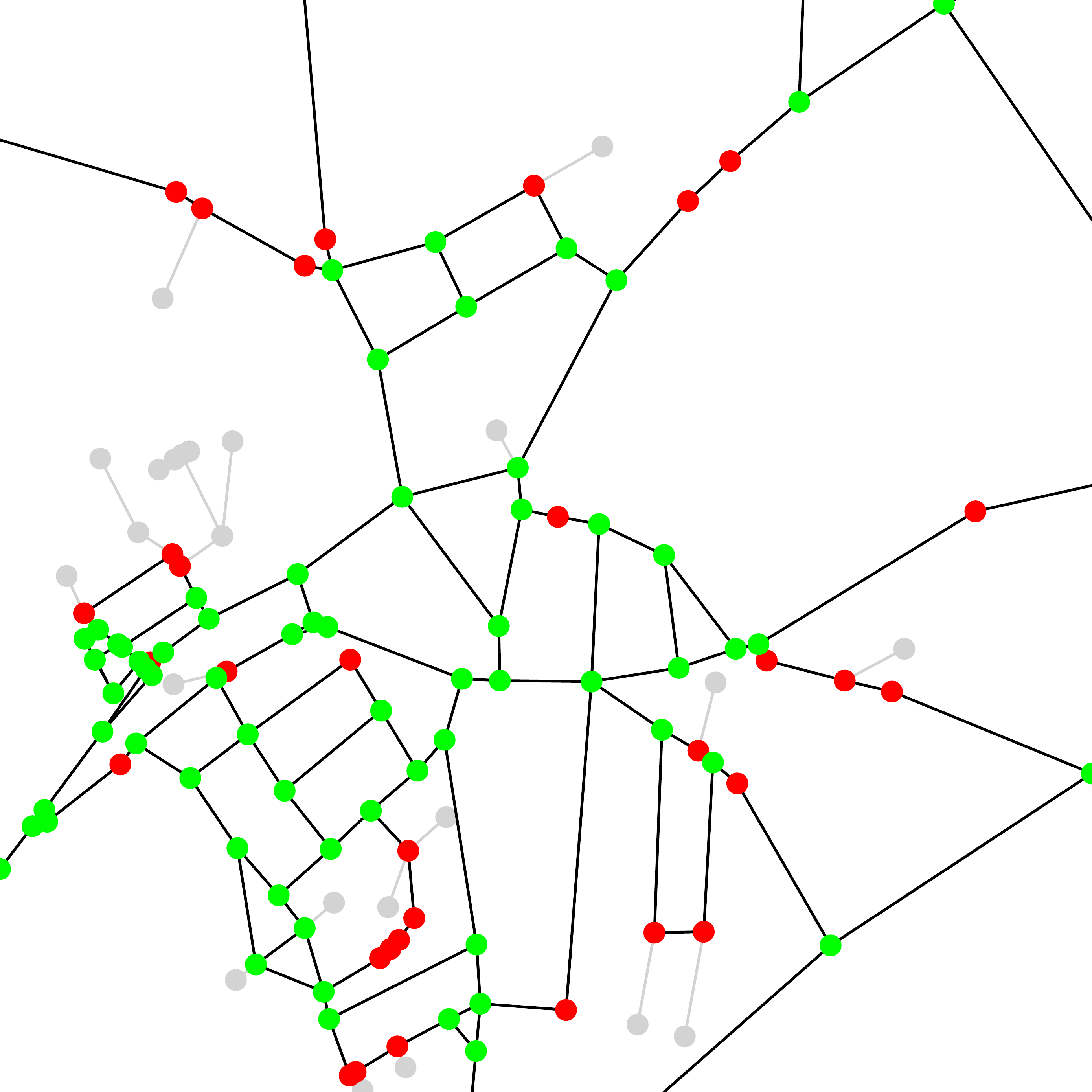}
\par\end{centering}

}
\par\end{centering}

\begin{centering}
\subfloat[OSM TopoCore]{\begin{centering}
\includegraphics[scale=0.137]{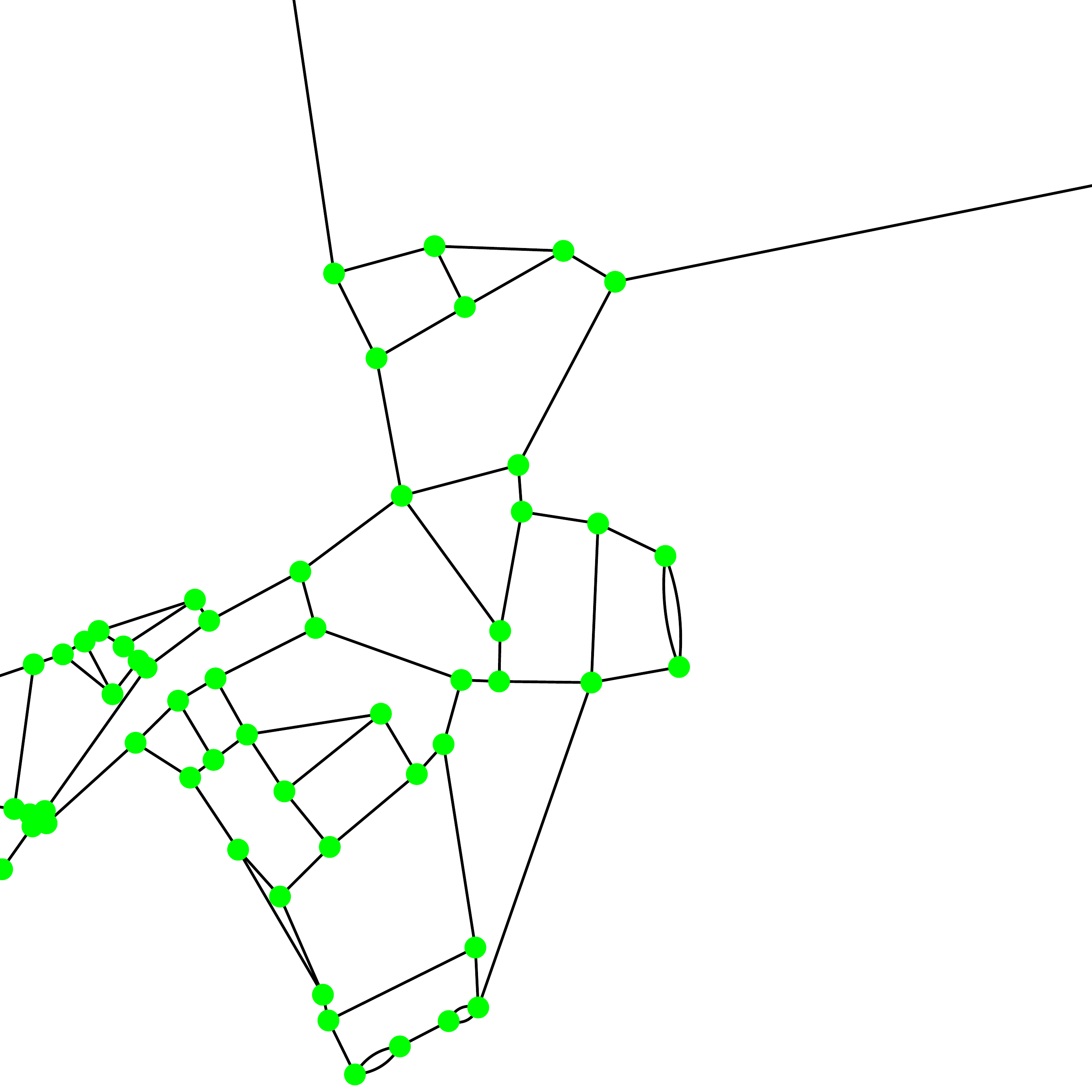}
\par\end{centering}

}\subfloat[DIMACS TopoCore]{\begin{centering}
\includegraphics[scale=0.137]{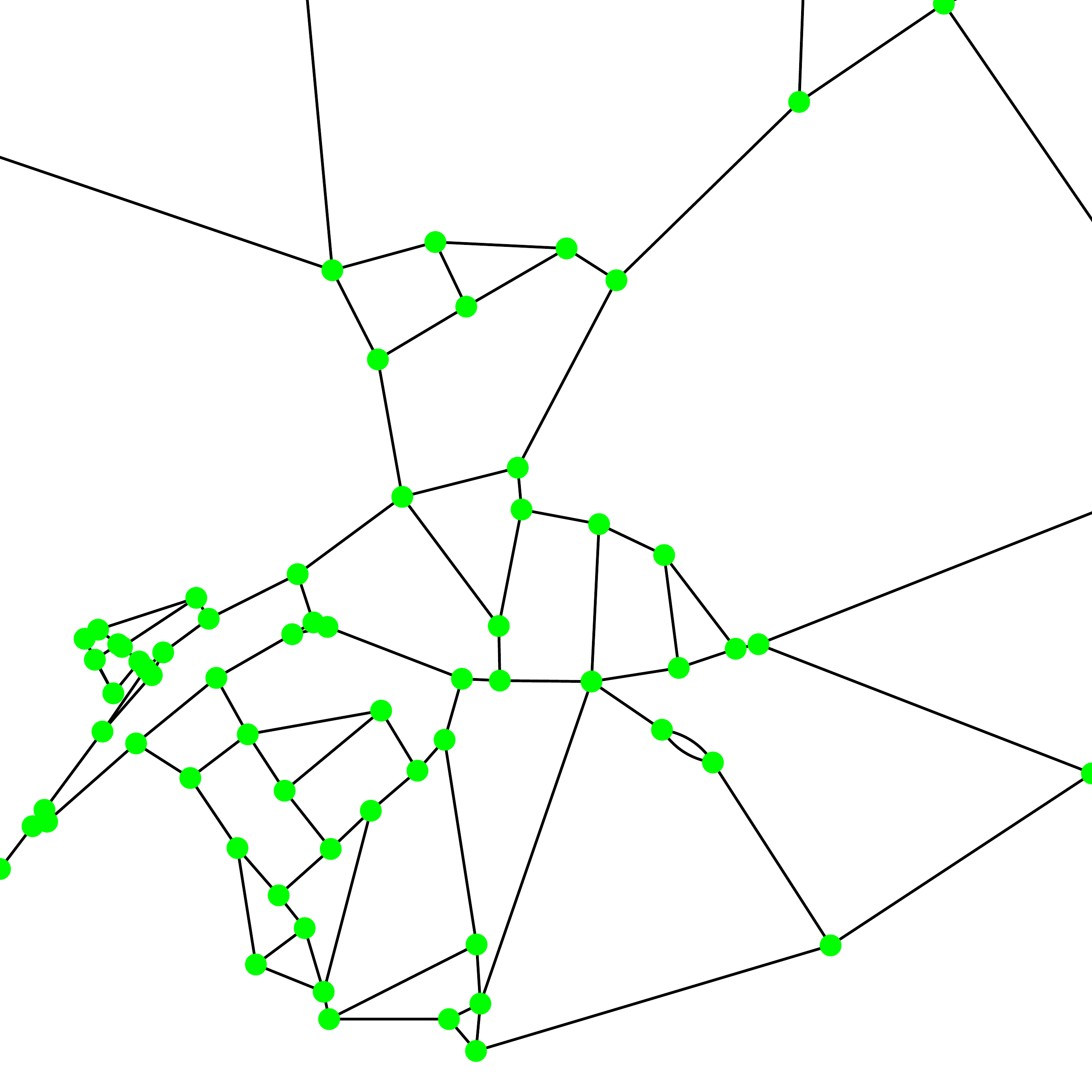}
\par\end{centering}

}
\par\end{centering}

\caption{\label{fig:topo-core}%
OSM (left) and DIMACS (right)
data sources of the area with a longitude in {[}8.50103, 8.52117{]} and latitude in {[}48.9476,48.9596{]}.
Nodes are drawn at geographical position. Arcs are drawn without direction for clarity.
Non-core nodes are red. Nodes not in the largest biconnected component are grayed out. Nodes in the TopoCore~(see Section~\ref{sec:core}) are green.}
\end{figure}

\section{Introduction}

Road networks of large geographic regions such as Europe or the U.S.\
easily consist of hundreds of millions of nodes, and collaborative spatial
data collection efforts, such as OpenStreetMap~(OSM)~\cite{osm},
have seen growths in node size by two orders of magnitude over the
last years. On such large networks, Dijkstra's classical shortest
path algorithm~\cite{d-ntpcg-59} incurs substantial running times
of several seconds even on modern computer hardware. This is too slow
for many applications such as navigation, route planning, location-based
services, range and trajectory queries, k-nearest-neighbor search,
and other queries on spatial network databases. Hence, the past decade
has seen numerous research (by both theoretical and applied
communities) into techniques that accelerate shortest path queries.
For an overview see the recent surveys~\cite{bdgmpsww-rptn-14,s-spqsn-14}. 

Assuming that the graph metric is fixed or does not change too often, these techniques offer very fast queries at  considerate preprocessing effort, enabling route planning services that serve millions of users per day. 
However, if instead costs change for every query, these techniques cease to provide benefit over Dijkstra's algorithm. 
Yet, in practice, even \emph{the same user} might prefer a quickest route in the morning but a safe and fuel-efficient route back home.

This scenario is considered in Personalized Route Planning~(PRP), a problem that was recently
introduced in a VLDB best paper~\cite{fns-opca-14}. Here, every
arc in the road graph is associated with a vector $\arccost$ of several
non-negative numeric costs such as for example travel time, distance, speed,
emissions, and energy consumption. The input of a query, in addition
to the source and the target node, consists of a cost vector $\queryweight$ with non-negative
entries. In the search, every arc is associated with the scalar
product of $\queryweight$ and~$\arccost$. The output consists of the shortest path 
with respect to this weighted sum of costs. Solving the PRP problem efficiently seems very
useful in order to construct route planning services that adapt
to the individual needs of every person. 

Unfortunately, in practice not all routing constraints can be modeled as a linear combination of additive costs. 
For example, summing up height limitations is not meaningful (\ie, a 3\,m high truck will not fit through two consecutive tunnels of 2\,m height).
A similar observation holds for vehicle weight limitations or the limit on the maximum slope that a vehicle can climb. 
Further constraints are
the avoidance of certain road categories, such as for example highways, city centers, or water conservation zones (which trucks with dangerous goods are not allowed to traverse).
In this work, we generalize PRP to also support such restrictions.

\subsection{Related Work}

The classic solution to solving shortest path problems on road networks is Dijkstra's algorithm~\cite{d-ntpcg-59}. 
Slightly faster queries are achieved by employing bidirectional search from both source and target~\cite{d-lpe-62,p-bs-71}. %
Furthermore, A* (or \emph{heuristic}) search~\cite{hnr-afbhd-68,p-bs-71} using easily available bounds (e.g., euclidean distance) is still a common choice. 
However, some studies, such as~\cite{gh-cspas-05}, have come to the conclusion that on road networks, A* with euclidean distance bounds is not necessarily beneficial over Dijkstra's algorithm; it can even slightly decrease efficiency. 
We have witnessed similar behavior in preliminary experiments in our specific setting.%

Many techniques have been proposed for further acceleration. Nearly
all of these divide the work into two phases: In a \emph{preprocessing
phase} the graph is augmented with auxiliary data that is then exploited
during the \emph{query phase} for faster shortest path or distance
retrieval. A good overview of techniques is given in~\cite{bdgmpsww-rptn-14,s-spqsn-14}.
Examples are graph \emph{partition}-based techniques~\cite{dw-cpiqr-13,epv-edmss-11,sww-daola-00},  %
landmark-based A{*}~(ALT)~\cite{ep-olbrp-13,gh-cspas-05},
\emph{Contraction Hierarchies}~\cite{gssv-erlrn-12,m-o-12}, %
and \emph{Hub Labeling}~\cite{DBLP:conf/esa/DellingGPW14}, %
the latter of which can be implemented on a DBMS~\cite{adfgw-hldbl-12}.

Above techniques work on the common assumption that costs
are known during the preprocessing phase. Since the preprocessing
effort is substantial, %
this can have a deterrent effect for real applications. 
Hence, techniques have been proposed that further subdivide the preprocessing phase, resulting in tool chains
that relatively quickly \emph{customize} the preprocessing to new costs~\cite{dgpw-crprn-13,dsw-cch-sea-14}, 
boiling them down to a singled fixed scalar cost to be considered by queries. 
Employing heavy parallelization on multi-core machines---or even multiple GPUs~\cite{dkw-cddgp-14}---these techniques achieve customization times faster than a single Dijkstra query. However, if costs change for \emph{every} query, spending so much computational effort seems questionable.\footnote{in a server setting, such resources could serve other clients in parallel; in a client setting they might not be available}
This is the case
for the scenario considered in our work: Personalized Route
Planning~(PRP), introduced by~\cite{fns-opca-14}, where it is approached based on \emph{$k$-path covers}.
A $k$-path cover $C$ is a small node subset of the original graph such that any simple (loop-free) path contains at most $k-1$ successive nodes that are not in~$C$. 
The core idea for accelerating PRP queries consists of computing a coarsened path that only contains nodes in $C$ where possible. 
Unfortunately, computing a minimum $k$-path cover is NP-hard~\cite{DBLP:journals/dam/BresarKKS11}.
For this reason in~\cite{fns-opca-14} approximate solutions were used.
Note that the $k$-path cover approach is inspired by the \emph{$k$-skip
covers} introduced in~\cite{tsp-osp-11}. The main difference is that
$k$-skip covers only guarantee that any \emph{shortest }(i.e., w.r.t.~a
fixed scalar cost) path contains at most $k-1$ successive nodes not in
$C$. The concept of $k$-skip covers is related to \emph{shortest path covers}~\cite{adfgw-h-13}, %
which have been used to show worst-case bounds for many speedup techniques (on graphs with small shortest path cover size).

The PRP problem is essentially a high-dimensional, linear multi-criteria search problem, related to the \emph{parametric shortest path problem}. %
Extensions of known preprocessing techniques to multi-criteria optimization have been proposed, but were only evaluated experimentally for the bi-criteria~\cite{gks-rpfof-10}
and tri-criteria~\cite{fs-pcchm-13} case. Even for the three criteria
of travel time, travel distance, and fuel consumption (which are even
quite correlated), diminishing returns in terms of query speed over preprocessing effort have been reported~\cite{fs-pcchm-13}.
Related approaches include Pareto-SHARC~\cite{dw-pps-09}, which drops exactness in its practical variant, and
Contraction Hierarchies with edge restrictions~\cite{DBLP:journals/jea/GeisbergerRST12}. %

\subsection{Our Contribution}

The primary results of our work are:

\begin{itemize}

\item We generalize Personalized Route Planning~(PRP) to support a more rich set of restrictions. The generalization allows to model, for example, maximum vehicle heights (\eg, for tunnels) and maximum vehicle weights (\eg, for bridges) as well as user-preferences such as avoidance of highways. 

\item A new preprocessing-based algorithm for PRP, extending the bilevel Dijkstra of~\cite{sww-daola-00}.
While we build on basic and easy to implement concepts, in combination our approach is better at~PRP than the state-of-the-art. %
A key ingredient is efficient identification of topologically important core nodes, 
while preserving all (not just shortest) paths. 
Figure~\ref{fig:topo-core} shows aspects of our construction, which is computed optimally in time linear in the size of the input graph. 

\item We conduct an extensive experimental study on a large set of real-world road graphs of different data sources. 
Our algorithms achieve significantly faster personalized route planning queries than previous approaches at less preprocessing costs.
 Furthermore, our query times are well below one second even on the largest instance tested for random long-distance queries. This is fast enough for a wide range of applications. Note that in practice most queries are short-distance that result in even lower query times.

\item Our analysis further shows that performance gains significantly vary depending on the data source---as opposed to just the geographical instance considered. 
While observed before, overall 
it is surprisingly underreported in the literature on route planning in road networks. 
We conclude that ranking road networks just by node count is not meaningful, and cross comparisons of the performance of route planning techniques are inconclusive without careful consideration of the respective data sources used for experimental evaluation.

\end{itemize}

\subsection{Outline}

We start with basic notation in Section~\ref{sec:prelim}.
In Section~\ref{sec:costs}, we formalize generalized arc costs supported by our approach.
Section~\ref{sec:node-order} discusses fine-tuning
Dijkstra's algorithm, since it is a central search subroutine for our query algorithm. 
In Section~\ref{sec:Bilevel-Dijkstra}, we describe how Dijkstra's algorithm is adjusted to make use of our preprocessing scheme. %
In Section~\ref{sec:core}, we explain in detail how to precompute the TopoCore and the TopoCore-IS. 
Finally in Section~\ref{sec:Experiments}, we report
methodology, setup and results of our careful experimental evaluation.

\section{Preliminaries}\label{sec:prelim}

We denote by $G=(V,A)$ a \emph{directed graph} with \emph{node set}~$V$ and \emph{arc set} $A\subseteq V\times V$. 
An \emph{undirected graph} is denoted by $G=(V,E)$ where $E$ is the \emph{edge set}.
 For road networks,
a node corresponds to a position on the earth's surface and an arc
to a road segment between two positions. In particular, \emph{not} every node
models a road intersection. For most arcs $(u,v)$ there is a \emph{back-arc}
$(v,u)$. However, there are notable exceptions such as one-way streets
or highways, which are modeled as two separate one-way streets. 
We consider multi-cost graphs, where each arc is associated with several costs, such as travel time
or distance. Denote by $k$ the number of costs. Formally, we have
a function $\arccost:A\rightarrow\mathbb{R}_{\ge0}^{k}$.
An \emph{$st$-path}  between a source node $s$ and a target node $t$, is a sequence $su\ldots vt$ of pairwise adjacent nodes.
A graph is called \emph{biconnected} if, after removing any node $v\in V$, the remaining graph $G-v$ is still connected.
A \emph{biconnected component}~(BCC) is a subgraph of $G$ that
is biconnected. An independent set $I$ is a subset of $V$ such that
no two nodes $u,v\in I$ are incident, \ie, no edge $\{u,v\}\in E$ exists.

\section{Generalized Costs}\label{sec:costs}

In its original formulation~\cite{fns-opca-14}, the PRP problem  consists of finding a path of minimum user-specified linear combination of \emph{additive} costs.
However, this is too restrictive in practice as some important constraints cannot be modeled as additive costs.
For example, one cannot simply add height limitations of two consecutive tunnels.
Other real-world restrictions such as vehicle width, vehicle weight, or maximum climbing ability (depending on the slope) essentially fall into the same category: Every road has a certain threshold value (\ie, the tunnel height), and if the vehicle's characteristic value (\ie, its height) is above this threshold, the vehicle is not allowed to traverse the road.
Clearly, adding these threshold values is not meaningful, instead one needs to compute the minimum of thresholds: A vehicle can pass through every tunnel on a path, if and only if it can pass through the lowest tunnel.
Restrictions that are formalized by upper bounds on vehicle characteristics are the most common.
However, there also restrictions that result in a lower bound. An example is the minimum required speed on highways that bans vehicles that cannot go fast enough.

Another source of restrictions is that some road categories are forbidden for some vehicle types. 
For example many city centers ban large trucks. Some trucks carry dangerous goods and are therefore not allowed in water conservation zones. Some drivers want to avoid highways with toll.
All of these restrictions have in common that some roads are flagged and some vehicles are not allowed to traverse them.
It is possible to regard them as 1-bit height-limitations. 
However, we prefer another view: We attach to every road a bitfield where the $i$-th bit stands for the $i$-th restriction of this type.
By convention we say that a bit being set means that a road can be traversed.
A path can be traversed if every road in it can be traversed.
Formally this consists of computing the bitwise-and of all road bitfields and testing the bits in the result against the vehicle restrictions or user preferences.

We support all these criteria by generalizing the PRP scenario.
The user does not input a vector of query weights~\queryweight, but an arbitrary function $f$ that fulfills a set of requirements.
We require $f$ to map cost vectors onto a value from $\mathbb{R}_{\ge 0} \cup \{\infty\}$. 
We further need an operation $\circ$ that combines two cost vectors. We require that $\circ$ is associative, i.e., for any cost vectors $\arccost_1$, $\arccost_2$, and $\arccost_3$ we require that $(\arccost_1 \circ \arccost_2) \circ \arccost_3 = \arccost_1 \circ (\arccost_2 \circ \arccost_3)$.  
Furthermore, it must not matter whether we first combine two cost vectors $\arccost_1$ and $\arccost_2$ and then apply $f$, or whether we first apply $f$ to both vectors and then compute the sum of the results. 
Formally, we require that $f(\arccost_1 \circ \arccost_2) = f(\arccost_1) + f(\arccost_2)$, which is the definition of $f$ being a semigroup homomorphism.

In the case of linear combinations, $f$ is the scalar product with $\queryweight$, and the $\circ$-operation is the component-wise addition. 
However, we can also do component-wise minimum or maximum, since it is associative, and even choose different operations for different cost components.
The right operation for height limitations (and similar restrictions), is to compute the minimum of all height limitations. 
The function $f$ then maps the cost vector onto $\infty$ if the vehicle is too high and otherwise ignores that cost component. 
The $\circ$ operation for road categories is the bitwise-and operation, which is fortunately also associative.
The function $f$ tests whether a certain bit, such as the highway bit, is set or not. 
Depending on the outcome $f$ evaluates to $\infty$ or $f$ looks only at the other cost components. 

\section{Tuning Dijkstra's Algorithm}\label{sec:node-order}

Dijkstra's algorithm \cite{d-ntpcg-59} is the textbook solution to
the shortest path problem, and many modern techniques still use it as a subroutine. Fine-tuning its implementation therefore directly results in better overall running times, but it also tightens the baseline for reporting speedups. (The speedup of a technique, which is used as an indication of machine-independent performance, is measured in terms of its query speed in relation to an implementation of Dijkstra's algorithm.) 
See, \eg, \cite{bdgmpsww-rptn-14} for a detailed discussion.

To ensure reproducibility of our experimental findings, we document details of our implementation and the reasoning behind the choices we made, as much as space allows. As datastructures we use an adjacency array representation of the graph and a 4-ary heap as queue, %
see~\cite{bdgmpsww-rptn-14} for details.

\subsection{Node Orders}

Node data is usually stored as a large array and the node-IDs correspond
to the offset in this array. A small ID-difference therefore implies
a high likelihood that the data of both nodes is loaded simultaneously into the cache. Dijkstra's
algorithm works by accessing the memory attached to the two endpoints
of an arc directly after another. If both are in cache, memory access times decreases. To illustrate this influence
we consider three node orders as in~\cite{dgnw-phast-13}: (a)~random order, (b)~input order, and
(c)~DFS pre-order. A random order performs the worst as it does not have much locality. The quality of the input order solely depends
on the data source. Usually it has some locality as nodes often appear
in the order that they were added to the dataset and adjacent nodes
are often added successively. The DFS pre-order consists of picking
a random root node and running a depth-first search. Nodes get ordered
in the way they are first visited. Every node with pre-order ID $i$
that is not the root or a leaf in the tree (i.e. the vast majority
of the nodes) will have two neighbors with directly adjacent node
IDs: The parent node has ID $i-1$ and the first child has ID $i+1$.
This covers most arcs as in road networks most nodes have degree 3 or less.

\subsection{Bidirectional Dijkstra}

Dijkstra's algorithm works by visiting all nodes around the source
node increasing by distance until the target node is reached. A speedup
can be gained by visiting the nodes around the source and the target
node simultaneously. This variant is called \emph{bidirectional} and
was first described in~\cite{d-lpe-62}. The central idea consists
of running two instances of Dijkstra's unidirectional algorithm simultaneously.
The first search explores the nodes close to the source node, while
the other explores the nodes around the target node. Once a node is
reached by both searches, a (not necessarily shortest) path is found.
Denote by $\mu$ the length of the shortest path found so far. Further
denote by $d_{F}$ the distance of the next node in the forward instance's
queue and by $d_{B}$ the distance of the next node of the backward
instance. We abort the search once $d_{F}+d_{B}\ge\mu$, as any path
that we find from that point on, has a distance of at least
$\mu$. Several \emph{alternation strategies} exist that decide from
which of the two queues a node should be popped and processed~\cite{p-bs-71,ww-sutsp-07}:
The strategy \emph{alternation} (alt) switches each step between forward
and backward search. The \emph{min-key} strategy (mk) picks the forward
search if $d_{F}\le d_{B}$. The \emph{min-queue-size} strategy (mq)
picks the forward search if the backward queue size is not smaller
than the forward queue size. 
Note that if the considered graph is directed, the backward search must operate on the reversed graph instead of the input graph. 

\section{Bilevel Variant of Dijkstra's Algorithm}\label{sec:Bilevel-Dijkstra}

A bilevel Dijkstra is a preprocessing-based technique to accelerate
shortest path queries. It is a variant of the technique introduced
in \cite{sww-daola-00}. In the preprocessing phase a \emph{core graph}
$G_{C}=(V_{C},A_{C})$ is computed. Think of this core graph as a
coarsened subgraph containing all major roads. The query phase is a bidirectional
variant of Dijkstra's algorithm. Conceptually, it first searches locally around
the source and the target nodes until the core is reached on both
sides. From there on the search is restricted to the core graph.
This decreases query times because $G_{C}$ is smaller than $G$ and
therefore only parts of the graph have to be searched.

Formally the nodes $V_{C}$ of $G_{C}$ are a subset of $V$ and called
\emph{core nodes}. Determining the right set of core nodes is crucial
for performance and detailed in the next Section~\ref{sec:core}.
The arcs of the core are defined as following: For every loop-free path $v_{1}v_{2}\ldots v_{k}$
for which only the endpoints $v_1$ and $v_k$ are in $V_{C}$ and all intermediate nodes
are in $V\backslash V_{C}$, there exists a \emph{shortcut} arc $(v_{1},v_{k})\in A_{C}$
in the core graph. Note that it is possible that multi-arcs are created
by this construction. The cost vector~$\arccost(v_{1},v_{k})$ of the shortcut is defined
as the combination of the cost vectors of the arcs within
the path, i.e., $\arccost(v_{1},v_{k})= \arccost(v_1,v_2)\circ \ldots \circ \arccost(v_{k-1},v_{k})$.

Given a core graph we compute a forward and a backward search graph
as follows: The forward graph $G_{F}$ is the union of $G$ and
$G_{C}$ without the arcs $(u,v)$ that leave the core, i.e., $u\in V_{C}$
and $v\in V\backslash V_{C}$. The backward graph $G_{B}$ is constructed
analogously: First compute the union of $G$ and $G_{C}$,
then reverse the direction of every arc and finally remove the arcs
leaving the core.

The query phase is a bidirectional variant of Dijkstra's algorithm.
The forward search is run on $G_{F}$ while the backward search runs
on $G_{B}$. We abort the search if $d_{F}+d_{B}\ge\mu$, where $\mu$
is the tentative distance, \emph{and} no queue contains a non-core node.

\section{Computing the Core Nodes}\label{sec:core}

\begin{figure}

\begin{centering}
\subfloat[OSM TopoCore-IS]{\begin{centering}
\includegraphics[scale=0.137]{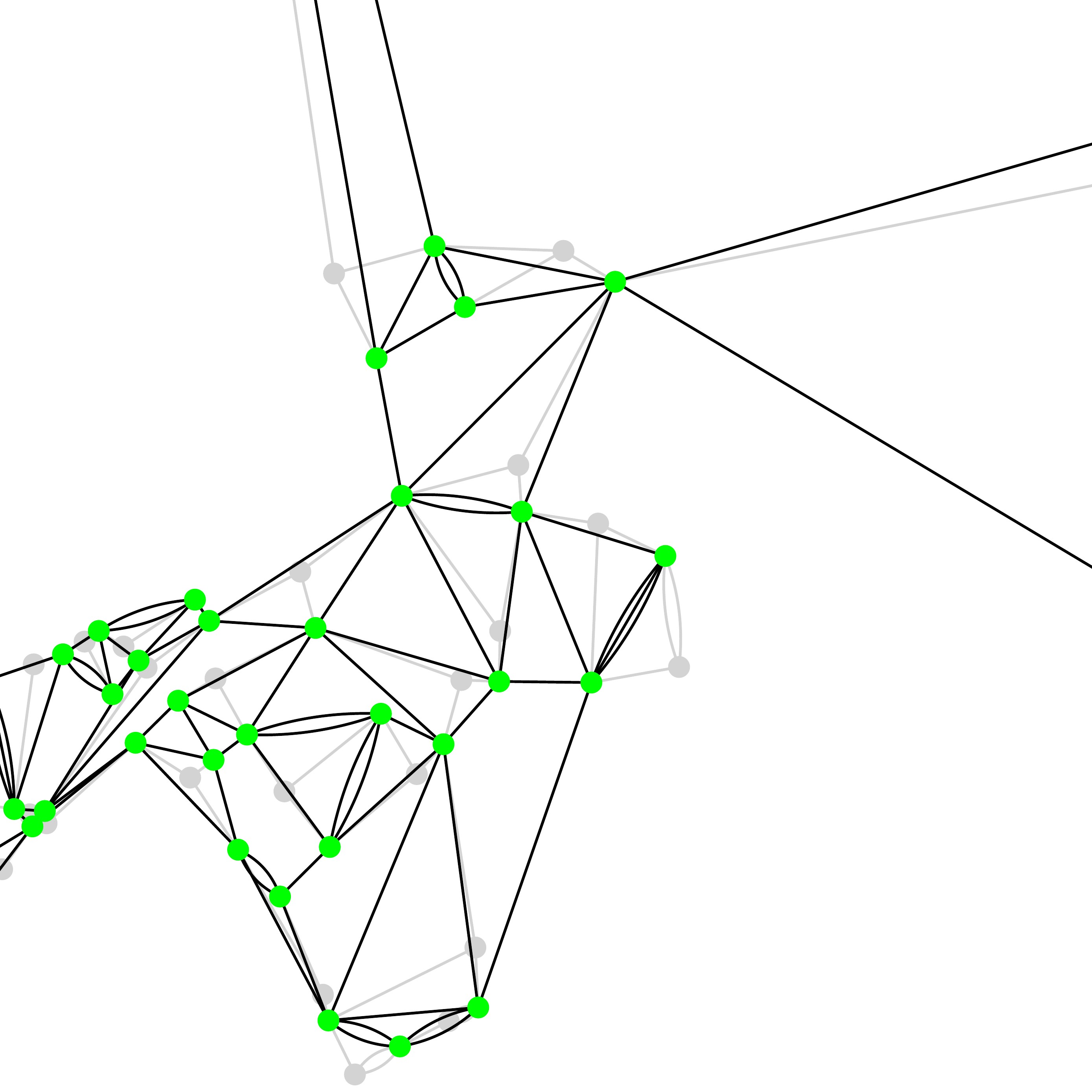}
\par\end{centering}

}\subfloat[DIMACS TopoCore-IS]{\begin{centering}
\includegraphics[scale=0.137]{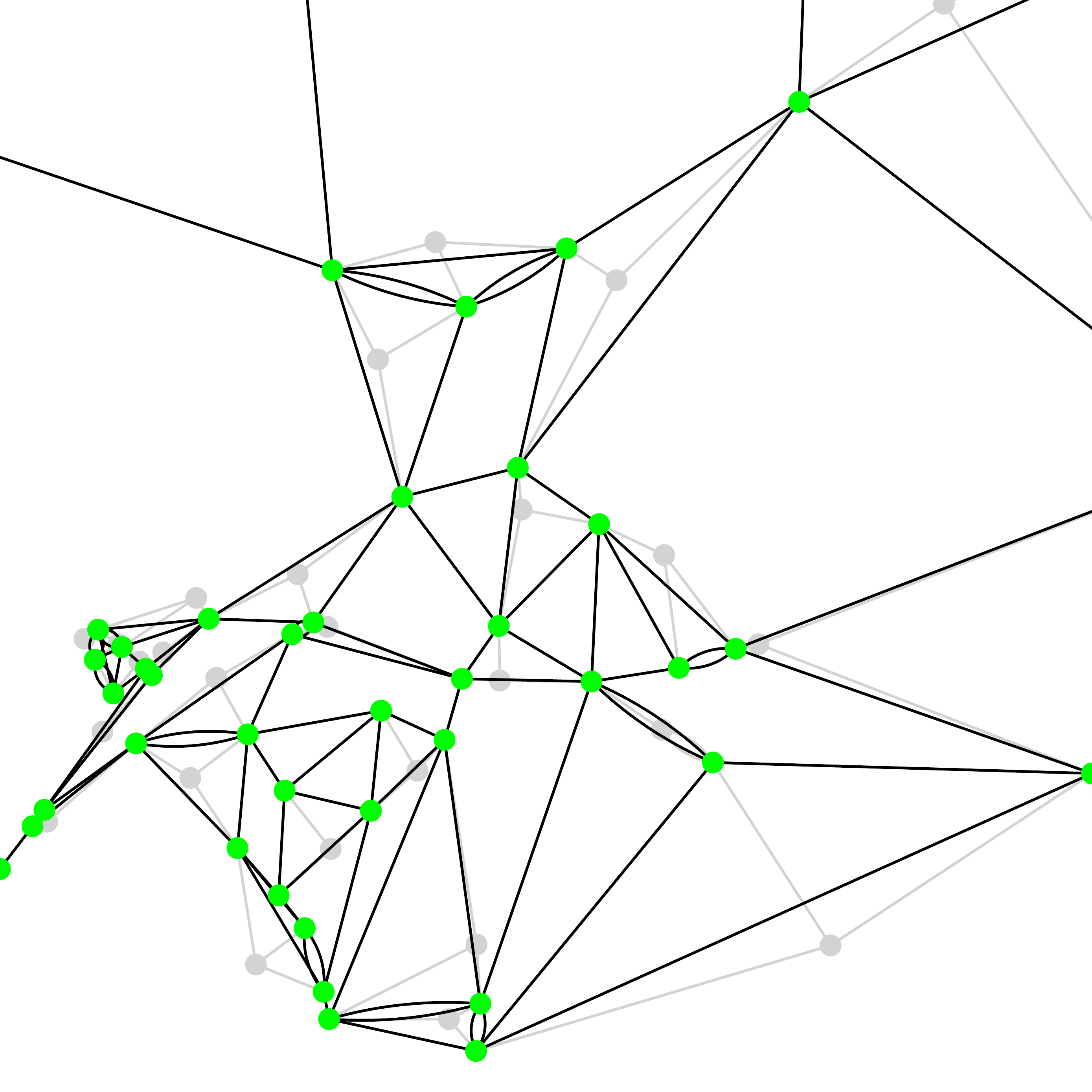}
\par\end{centering}

}
\par\end{centering}

\caption{
OSM (left) and DIMACS (right) data sources, \cf~Figure~\ref{fig:topo-core}.
The TopoCore-IS is drawn upon a grayed-out TopoCore, with added shortcuts between green nodes.
}
\label{fig:topo-core-is}
\end{figure}

In the previous section we described how a set of ``good'' core nodes is
used to realize a bilevel variant of Dijkstra's algorithm. In this
section we describe how to compute this set of ``good'' core nodes.
Initially, all nodes are core nodes. 
Then, for each node removed from the core, we potentially have to add shortcuts between \emph{all} pairs of neighbors, in order to maintain shortest path distances for the yet unknown objective function (to be specified in the query). Note that, unlike~\cite{dsw-cch-sea-14}, we must create multi-arcs if an original arc between two neighbors is already present (since we cannot tie-break for an unknown objective function). As the performance of Dijkstra's algorithm (and its bilevel variant) depends on both the number of nodes \emph{and} arcs, we would eventually experience diminishing returns if adding too many new arcs while removing nodes from the core.

Hence, our goal is to select as few core nodes as possible while restricting growth in the number of core arcs.
In the following, we describe three steps performed in succession to remove
nodes from the core, reducing its size and thus accelerating shortest
path queries. We refer to the core that is produced after Step 2 as
\emph{TopoCore}. The name was chosen to reflect that we exploit only
topological graph features. After Step 3, we refer to the core as \emph{TopoCore-IS}, where IS stands for independent set.

\subsection{Step 1: Removing Dead-Ends}
First, we compute the biconnected components of the input graph, employing a linear-time algorithm by Tarjan~\cite{t-dfslg2-72}. (For this, we ignore arc directions.)
Each dead-end like structure is its own tiny component. 
All that entails significant routing decisions, forms a single large component.
Hence, we keep every node in the core that is contained in the largest biconnected component.
Note that we do not add any shortcuts in this step. 

\subsection{Step 2: Removing Chains}

Consider the graph induced by all core nodes. 
Note that removing a node with only two neighbors from the core, while adding shortcuts between its neighbors, does not increase core arc size.
Better yet, in our inputs, such nodes are often not isolated but form chains between two nodes of higher degree. 
Moreover, these chains may grow by first applying Step~1, as intersections exist, where all but two roads lead to dead-ends. 
First removing dead-ends turns such intersections into degree 2 nodes.
We identify such chains and add shortcut arcs to the core that bypass them, removing bypassed nodes from the core.
Note that the resulting \emph{TopoCore} may contain multi-arcs. See Figure~\ref{fig:topo-core} for an illustration. 

\subsection{Step 3: Removing Degree-3 Nodes}
Ideally, we would like to remove even more nodes from the core.
In case of undirected simple graphs, removing a node of degree~$d$ (\ie, with $d$ neighbors) from the core removes $d$ edges (to these neighbors) from the core, while adding $d(d-1) / 2$ new edges to the core, \ie, a net increase of $d(d-3) / 2$. Hence for $d=3$, the number of edges in the core remains unchanged but the number of
nodes decreases. It is therefore beneficial to remove degree-3 nodes
from the core for a reduction in queue operations during search. Our experiments in Section~\ref{sec:Experiments} show, that there is an abundance of degree-3 nodes in the TopoCore.

In reality, our input graphs are directed and Step~2 may have created multi-arcs.
We deal with multi-arcs by defining the node degree as the number of incident arcs.
Furthermore, for directed graphs, removing a high-degree node might not necessarily result in a net increase of core arcs. (For example, consider a node with a single in-arc: Regardless of its out-degree, removing the node from the core would decrease the number of arcs in the core by~1.) Since road networks are mostly undirected (\ie, most road segments can be traversed in both directions), we do not try to exploit such cases, \ie, we ignore arc directions to determine node degrees. 

Hence, the idea is to remove degree-3 nodes from the core. But we cannot just remove all of them, as removing
a node may increase the degree of its neighbors, turning a degree-3
node into a higher degree node. Therefore, we first compute an independent
set of degree-3 core nodes (iterating over the nodes in DFS pre-order and greedily adding
degree-3 nodes to the set that have no adjacent degree-3 node in the set). We then remove only this independent set from the
core. See Figure~\ref{fig:topo-core-is} for an illustration of the resulting \emph{TopoCore-IS}.
One could try to apply this procedure iteratively, but our experiments indicate that in the TopoCore-IS only few degree-3 nodes remain.

\subsection{Node Orders}

The order in which node data appears in memory has, as argued
in Section~\ref{sec:node-order}, a significant impact on query speed. We first reorder the input graph using a DFS pre-order. We then compute the core and move
core nodes to the front of the order. This yields DFS pre-order
inside of the core. Outside of the core the nodes also have an order
that locally behaves DFS-like. The arcs bridging the largest
node-ID differences tend to be arcs entering or leaving the core. 

\section{Experiments} \label{sec:Experiments}

\begin{table}[b]
\caption{\label{tab:size}The sizes of our benchmark graphs. We report the number of nodes~$|V|$, the number of arcs~$|A|$, and the node degree distribution.}

\begin{tabular}{m{0.3\columnwidth}m{0.3\columnwidth}m{0.26\columnwidth}}
\toprule
 & $|V|$ & $|A|$ \tabularnewline
\midrule
OSM-BaWü & 3\,064K & 6\,184K \tabularnewline
OSM-Ger & 20\,690K & 41\,792K \tabularnewline
OSM-Eur & 173\,789K & 347\,997K \tabularnewline
DIMACS-Eur & 18\,010K & 42\,189K \tabularnewline
DIMACS-US & 23\,947K & 57\,709K \tabularnewline
\end{tabular}

\begin{tabular}{lrrrrr}
\toprule
 & \multicolumn{5}{c}{\#\,Nodes per degree}\tabularnewline
 & 1 & 2 & 3 & 4 & 5+\tabularnewline
\midrule
OSM-BaWü & 13.3\% & 72.6\% & 12.6\% & 1.2\% & 0.01\%\tabularnewline
OSM-Ger & 14.2\% & 70.9\% & 13.5\% & 1.3\% & 0.01\%\tabularnewline
OSM-Eur & 12.1\% & 76.7\% & 10.1\% & 1.1\% & 0.01\%\tabularnewline
DIMACS-Eur & 26.5\% & 18.7\% & 49.1\% & 5.7\% & 0.1\%\tabularnewline
DIMACS-US & 19.9\% & 30.3\% & 39.0\% & 10.7\% & 0.1\%\tabularnewline
\bottomrule
\end{tabular}

\end{table}

\subsection{Setup and Methodology}
We implemented our algorithms in C++, compiling on g++ 4.6.3 with optimization level \texttt{-O3}.
Our experiments were performed on a \emph{single core} of an Intel Xeon E5-2670 processor (Sandy Bridge architecture) clocked at 2.6\,GHz, with 64\,GiB of DDR3-1600 RAM clocked at 1.6\,GHz, 20\,MiB of L3 and 256\,KiB of L2 cache.

\begin{figure}[b!]
\begin{centering}
\subfloat[OSM-BaWü]{\begin{centering}
\includegraphics[scale=0.06]{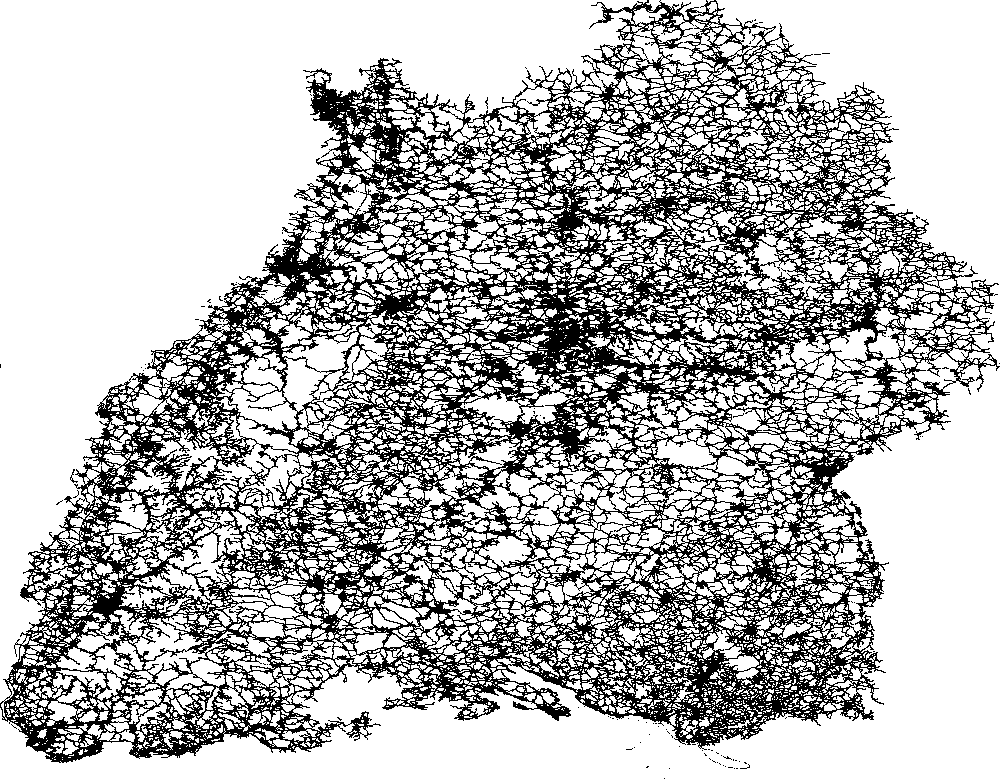}
\par\end{centering}

}~\subfloat[OSM-Ger]{\begin{centering}
\includegraphics[scale=0.11]{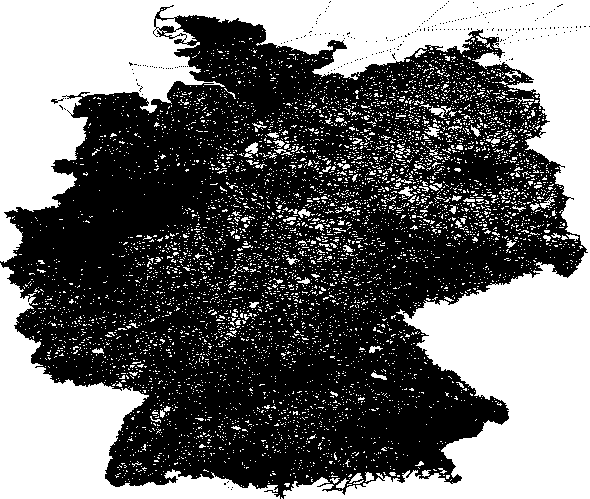}
\par\end{centering}

}~\begin{centering}
\subfloat[DIMACS-Eur]{\begin{centering}
\includegraphics[scale=0.084]{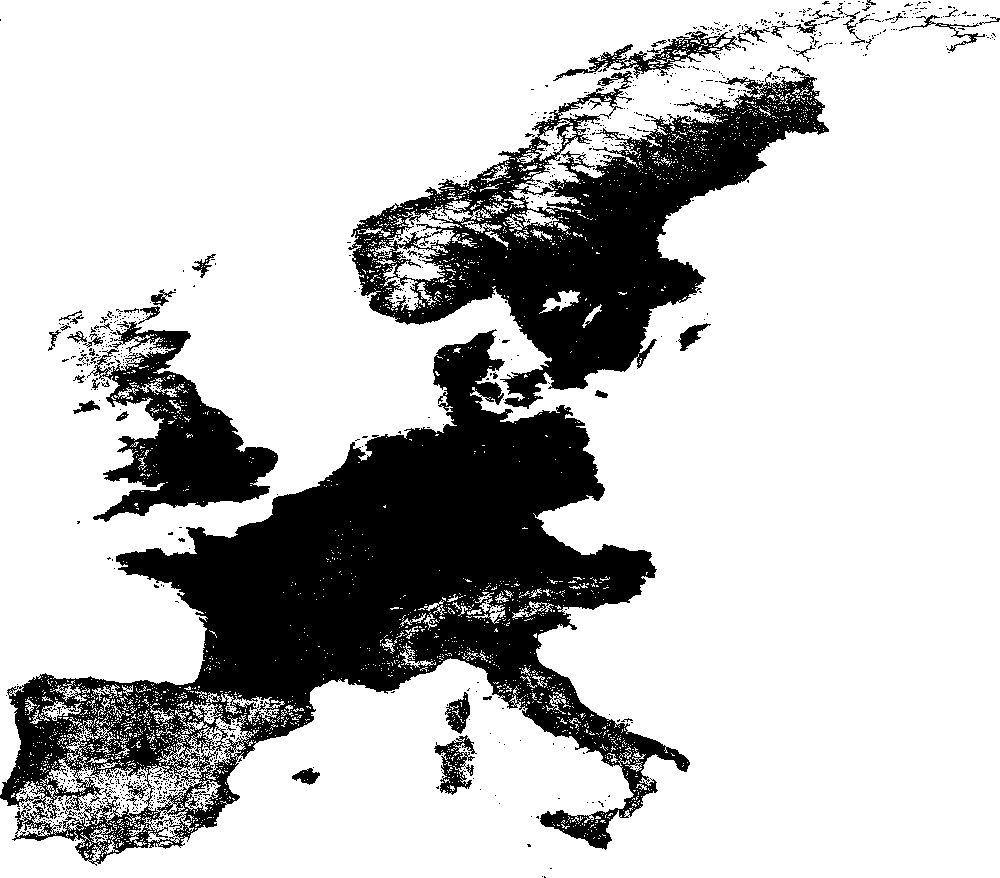}
\par\end{centering}

}
\par\end{centering}

\subfloat[DIMACS-US]{\begin{centering}
\includegraphics[scale=0.1]{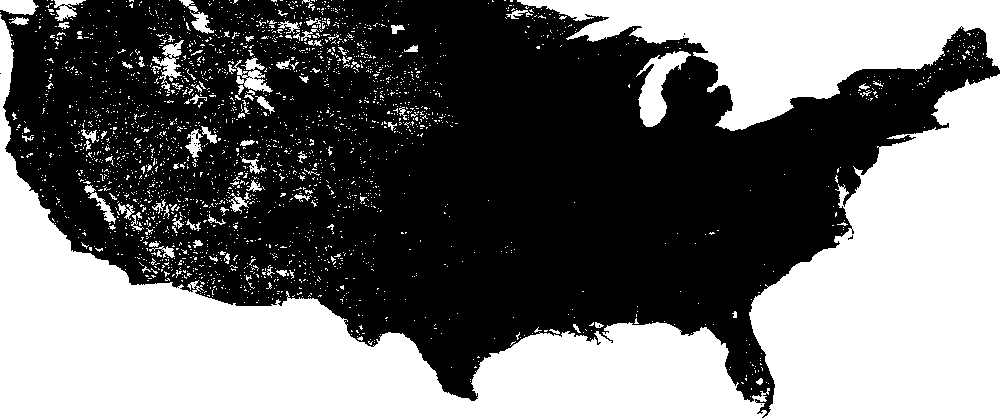}
\par\end{centering}

}~\subfloat[OSM-Eur]{\begin{centering}
\includegraphics[scale=0.12]{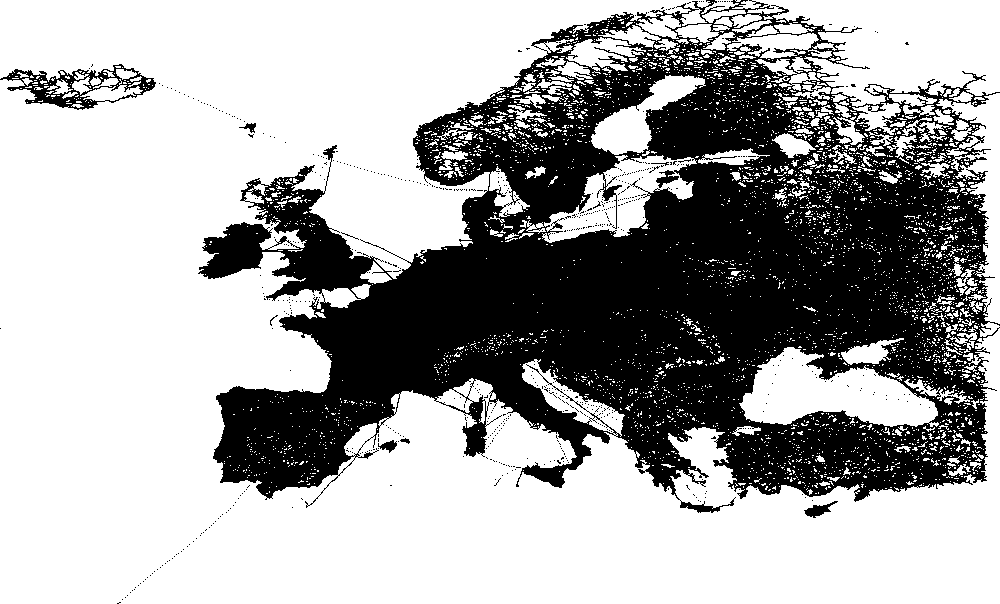}
\par\end{centering}

}
\par\end{centering}

\caption{\label{fig:graphs}The geographical regions corresponding to our benchmark graphs.}
\end{figure}

We use five different road networks of three different origins as our test instances. Table~\ref{tab:size} reports basic statistics. Figure~\ref{fig:graphs} depicts the geographical regions represented by the graphs.
The two DIMACS instances were published for the 9th DIMACS implementation Challenge~\cite{dgj-spndi-09}. 
DIMACS-Eur was compiled from NAVTEQ~\cite{navteq} data and kindly made available by PTV AG~\cite{ptv}, it includes the road networks of 17 Western European countries. 
DIMACS-US was derived from the \emph{UA Census 2000 TIGER/Line Files} produced by the Geography Division of the US Census Bureau.
The OSM instances were obtained from \url{http://download.geofabrik.de/} at 2014-10-23T20:22:02Z, courtesy of GeoFabrik GmbH~\cite{geofabrik}. From that data, we compiled our routing networks using the graph extraction tools provided by OSRM~\cite{luxen-vetter-2011} with the ``car'' profile. More precisely, we used this version of the code: \url{https://github.com/Project-OSRM/osrm-backend/tree/6f75d68d07a5d1a67219835a0638cd0a482a18f5}. 
OSM-BaWü is the road network of the state of Baden-Württemberg in Germany, OSM-Ger that of Germany.
OSM-Eur contains the road networks of 48 European regions, including western Russia.
We remove multi-arcs from the input and only keep the
largest strongly connected component to assure that between each pair
of nodes at least one shortest path exists. The numbers in Table~\ref{tab:size}
are the graph sizes after these standard cleanup procedures were applied.
Still, our OSM graphs are larger than those reported in \cite{fns-opca-14}; 
We suspect that the OSM data we use is more recent and therefore contains
more details. Note, however, that our graphs have a very similar average degree (which for a given data source, \ie, OSM in this case, indicates a similar degree distribution) and should therefore behave similarly. For future reference, we have made our OSM instances publicly available under~\url{http://i11www.iti.uni-karlsruhe.de/resources/roadgraphs.php} in the same format as used in the DIMACS challenge. The DIMACS instances are available under~\url{http://www.dis.uniroma1.it/challenge9/download.shtml}. 
 
We evaluate the performance of our algorithm with respect to the basic and the generalized PRP problem.
For the basic PRP problem we attach cost vectors with 8 entries to each
arc (as chosen for the largest graph evaluated
in~\cite{fns-opca-14}). Each cost entry is a 32-bit int. 
Each of the test instances provides travel time~$t$ for each road segment, and we infer a road distance~$d$ from the geographical positions of the segment end points.
Unfortunately, we do not have any further road metric that is available on every instance.
We therefore generate 6 further costs per arc: $100t/d$, $100d/t$, $100/d$, $100/t$, 1, and a random number between 0 and 100. 
Notice that none of these costs is a linear combination of the other costs.
We therefore have a sufficiently diverse structure to get meaningful results. 
For the generalized PRP we also have 8 costs but only the first 4 are additive. 
These are $t$, $d$, $100t/d$, and $100d/t$. 
The last 4 are thresholds such as needed for height limitations.
As we do not have real world data available we generate synthetic data.
For every arc $a$ and cost $c$ we throw a 1000-sided dice. 
If it lands on 0, we attach a random threshold between 0 and 100 to the cost $\arccost$ of the arc $a$. 
If the dice lands on any other number we assign a threshold of $+\infty$.
Note that we assign $+\infty$-thresholds with such high probability, in order to ensure connectivity of the graph. %

For all query time experiments we sampled 1000 uniform random source and target pairs. 
Note that uniform random queries are long-distance queries with high expectancy. Typically, most queries issued on real systems, e.g., navigation devices, are short-range queries and should be answered faster.
We make sure that queries are the same for different node orderings of the same graph (by permuting the pairs according
to the node ordering instead of picking a new independent set of 1000
random pairs). We further pick a query weight~$\queryweight$ of 8 random entries between 0 and
100 for each query. For the generalized PRP problem we interpret the last 4 entries as vehicle characteristics that must be below a threshold (such as for example the vehicle's height).
To avoid overflows all computations are done using 64-bit integer arithmetic. Our implementation of Dijkstra's algorithm stores 64-bit tentative distance values for each node.
It uses a 4-ary heap as queue.

\subsection{Preprocessing}

\begin{table}
\caption{\label{tab:prepro}Preprocessing time in seconds. ``BCC'' is the
time needed to compute the biconnected components. We also report
the time needed to randomly reorder all nodes and their incident arcs
and cost vectors in memory. }
\begin{centering}
\begin{tabular}{lrrr}
\toprule
 & Reorder Nodes & BCC & Insert Shortcuts\tabularnewline
\midrule 
OSM-BaWü & 1.2 & 0.8 & 0.7\tabularnewline
OSM-Ger & 22.8 & 6.7 & 5.8\tabularnewline
OSM-Eur & 304.0 & 150.1 & 202.8\tabularnewline
DIMACS-Eur & 22.1 & 7.2 & 6.5\tabularnewline
DIMACS-US & 25.3 & 9.1 & 8.3\tabularnewline
\bottomrule
\end{tabular}
\par\end{centering}
\end{table}

\begin{table*}
\caption{\label{tab:core-size}Core graph sizes. We also report the number of nodes and arcs of each core in percent of the input graph's number of nodes respectively arcs.}
\begin{centering}
\begin{tabular}{llrrrrrrr}
\toprule
 &  & Input & \multicolumn{2}{c}{BCC} & \multicolumn{2}{c}{TopoCore} & \multicolumn{2}{c}{TopoCore-IS}\tabularnewline
\midrule
OSM-BaWü & $|V|$ & 3\,064K & 2\,095K & 68.4\% & 270K & 8.8\% & 161K & 5.3\%\tabularnewline
 & $|A|$ & 6\,184K & 4\,489K & 72.6\% & 777K & 12.6\% & 730K & 11.8\%\tabularnewline
 \addlinespace
OSM-Ger & $|V|$ & 20\,690K & 14\,088K & 68.1\% & 1\,887K & 9.1\% & 1\,125K & 5.4\%\tabularnewline
 & $|A|$ & 41\,792K & 30\,267K & 72.4\% & 5\,430K & 13.0\% & 5\,088K & 12.2\%\tabularnewline
 \addlinespace
OSM-Eur & $|V|$ & 173\,789K & 116\,232K & 66.9\% & 13\,957K & 8.0\% & 8\,414K & 4.8\%\tabularnewline
 & $|A|$ & 347\,997K & 248\,209K & 71.3\% & 39\,145K & 11.2\% & 36\,789K & 10.6\%\tabularnewline
 \addlinespace
DIMACS-Eur & $|V|$ & 18\,010K & 11\,763K & 65.3\% & 7\,108K & 39.5\% & 4\,299K & 23.9\%\tabularnewline
 & $|A|$ & 42\,189K & 31\,584K & 74.9\% & 20\,347K & 48.2\% & 19\,387K & 46.0\%\tabularnewline
 \addlinespace
DIMACS-US & $|V|$ & 23\,947K & 16\,020K & 66.9\% & 7\,415K & 31.0\% & 4\,789K & 20.0\%\tabularnewline
 & $|A|$ & 57\,709K & 41\,412K & 71.8\% & 24\,201K & 41.9\% & 23\,754K & 41.2\%\tabularnewline
 \bottomrule
\end{tabular}
\par\end{centering}
\end{table*}

In Table~\ref{tab:prepro} we report the time needed by our preprocessing.
Computing the biconnected components and computing the shortcuts are
the most expensive algorithmic tasks. However, as the table shows,
its running time is dominated by seemingly unsophisticated operations
such as permuting all nodes in-memory. The reason is that the cost
vectors need a lot of space (32\,Byte per arc) and need to be reordered
as well. For example for OSM-Eur the arc cost data alone needs over
$|A|\cdot4\cdot8>$10\,GB of RAM. Shuffling memory is therefore a comparatively
expensive task. We therefore expect that in a productive implementation
the running time is not dominated by purely algorithmic aspects but parsing the input data should dominate.

Table~\ref{tab:core-size} details the sizes of the various obtained cores.
The first step of removing the nodes not in the largest biconnected
component decreases the node counts by roughly 30\% for all graphs.
How effective removing degree-2 nodes is depends on the graph. For
the OSM graphs core sizes decrease by a factor of 8 in terms of nodes.
The size decrease for DIMACS-US is only a factor 2 and for DIMACS-Eur
it is even only 40\% less nodes. Removing degree-3 nodes further decreases
the node count by 40\%. As expected the number of arcs does not decrease significantly in this final step.

Besides core sizes we also report in Table~\ref{tab:chain} the average
number of arcs in the degree-2 chains removed from the graph. A chain
is a sequence of at least 2 arcs where all intermediate nodes have
degree 2. Note that we first compute the biconnected components (BCC) before computing the
chains. This order increases the chain lengths increasing the effectiveness
of our technique. Again the numbers show that the OSM graphs have
more degree-2 nodes and thus longer chains. We further report the
number of degree-3 nodes. As expected this number significantly decreases
when going from TopoCore to TopoCore-IS.

\begin{table}
\caption{\label{tab:chain}The average number of arcs per degree-2 chain and
the remaining number of degree-3 nodes.}
\begin{centering}
\begin{tabular}{lrrr}
\toprule
 & Avg. \#\,arcs  & \multicolumn{2}{c}{Number of degree-3 nodes}\tabularnewline
 & per chain & TopoCore & TopoCore-IS\tabularnewline
\midrule 
OSM-BaWü & 7.2 & 249K & 20K\tabularnewline
OSM-Ger & 6.9 & 1\,738K & 137K\tabularnewline
OSM-Eur & 8.5 & 12\,741K & 1\,478K\tabularnewline
DIMACS-Eur & 2.7 & 6\,435K & 560K\tabularnewline
DIMACS-US & 3.2 & 5\,481K & 40K\tabularnewline
\bottomrule
\end{tabular}
\par\end{centering}
\end{table}

\paragraph{Memory Consumption}

\begin{table}
\caption{\label{tab:memory}Input graph size and additional memory needed by TopoCore and TopoCore-IS for $k=8$.}
\begin{centering}
\begin{tabular}{lrrr}
\toprule
Graph & Input & TopoCore & TopoCore-IS \\
\midrule
OSM-BaWü & 224MB & 28MB & 26MB \\
OSM-Ger & 1\,514MB & 194MB & 179MB \\
OSM-Eur & 12\,610MB & 1\,397MB & 1\,295MB \\
 \addlinespace
DIMACS-Eur & 1\,517MB & 726MB & 682MB \\
DIMACS-US & 2\,073MB & 859MB & 834MB \\
\bottomrule
\end{tabular}
\par\end{centering}
\end{table}

Suppose that the input graph has $n$ nodes and $m$ directed arcs and that the core graph has $n_c$ nodes and $m_c$ arcs. 
Further there are $k$ costs and each ID and cost entry is encoded using 32-bits. 
To store the structure of input graph in an adjacency array $4(n+1)+4m$ bytes are needed.
The cost vectors need another $4km$ bytes of storage.
The total space required by the input graph is thus $4((n+1)+(k+1)m)$.
Similarly the total additional space required by the core graph is $4((n_c+1)+(k+1)m_c)$. 
As we reorder all core nodes to the front, we do not need to explicitly store which nodes are core nodes but can compare the node ID to $n_c$.
Table~\ref{tab:memory} depicts the memory consumption for all benchmark graphs.

\subsection{Query}

Table~\ref{tab:dij-tuning} compares the performance
of Dijkstra's algorithm in its unidirectional
and bidirectional variants and with all three node orders.
Overall, bidirectional search with minimum-queue-size alternation
strategy yields the best query performance, consistently about 55\,\% faster than unidirectional search.
Additionally, DFS-reordered nodes improve query times by 19--23\,\%, compared to the input order.

However, we also note that the gap to unidirectional search on random order is much higher.
This raises the question of what is a good baseline for determining speedups of preprocessing techniques.
Especially if these techniques provide only comparatively low speedups (\eg, of one order of magnitude, because the considered scenario is so involved),
it is very important to carefully document the baseline.
While often undocumented, we believe that unidirectional search with input order is the variant
used in most other studies and therefore use it as baseline from here on, too.
(However, one could argue in favor of a random order, since it  
eliminates a dependency on the data source, which might or might not provide a good input order.)

In Table~\ref{tab:topo-core} we report the running times of our query algorithm on both variants of the PRP problem.
We observe that the running times are very similar for both problems. 
We conclude that the running time is bounded by the work done by Dijkstra's algorithm and not the time needed to evaluate the costs at the edges.
On graphs with an abundance of degree-2 nodes (such
as OSM) we achieve large speedups of approximately 30-55. On graphs
with fewer degree-2 nodes the results are less impressive but the
speedups of about 6.2-8.5 is still a significant improvement over the
baseline.

\begin{table}
\caption{\label{tab:dij-tuning}Query running time and number of queue-pop-operations for variants of Dijkstra's algorithm on the OSM-BaWü graph for the general PRP problem.
``random'', ``input'' and ``dfs'' are the node orders considered.
They vary in terms of running time because of cache-effects but not
in terms of pop-operations. ``mk'', ``alt'' and ``mq'' are the
alternation strategies. }
\begin{centering}
\begin{tabular}{lrrrr}
\toprule
 & \multicolumn{3}{c}{Time {[}ms{]}} & Nodes popped\tabularnewline
Dir & Random & Input & DFS & from queue\tabularnewline
\midrule 
uni & 470 & 265 & 223 & 1\,539K\tabularnewline
bi-mk & 371 & 216 & 176 & 1\,009K\tabularnewline
bi-alt & 343 & 188 & 156 & 938K\tabularnewline
bi-mq & 302 & 171 & 143 & 900K\tabularnewline
\bottomrule
\end{tabular}
\par\end{centering}

\vspace{0.5em}
\caption{\label{tab:topo-core}
Query running time (T) and number of queue-pop-operations (P) using the TopoCore~(TC) and TopoCore-IS~(TC-IS) techniques and speedup~(Sp.up) compared to an unidirectional baseline with input order. We use the min-queue-size alternation strategy.
	}
	\begin{center}
	\subfloat[Basic PRP Problem]{
		\begin{centering}
		\begin{tabular}{llrrrr}
		\toprule
		 &  & Input & TC & TC-IS & Sp.up\tabularnewline
		\midrule 
		OSM & T {[}ms{]} & 265 & 14 & 9 & 29.4\tabularnewline
		~~-BaWü& P {[}$\cdot10^{3}${]} & 1\,539 & 80 & 48 & 32.1\tabularnewline
		 \addlinespace
		OSM & T {[}ms{]} & 2\,914 & 118 & 80 & 36.4\tabularnewline
		~~-Ger & P {[}$\cdot10^{3}${]} & 10\,313 & 599 & 357 & 29.9\tabularnewline
		 \addlinespace
		OSM  & T {[}ms{]} & 32\,145 & 891 & 621 & 51.8\tabularnewline
		~~-Eur& P {[}$\cdot10^{3}${]} & 83\,938 & 3\,761 & 2\,266 & 37.0\tabularnewline
		 \addlinespace
		DIMACS & T {[}ms{]} & 1\,817 & 424 & 291 & 6.2\tabularnewline
		~~-Eur & P {[}$\cdot10^{3}${]} & 9\,015 & 1\,976 & 1\,195 & 7.5\tabularnewline
		 \addlinespace
		DIMACS & T {[}ms{]} & 3\,045 & 523 & 381 & 8.0\tabularnewline
		 ~~-US & P {[}$\cdot10^{3}${]} & 11\,912 & 2\,339 & 1\,513 & 7.9\tabularnewline
		\bottomrule
		\end{tabular}
		\par\end{centering}
	}
	\\
	\subfloat[Generalized PRP Problem]{
		\begin{centering}
		\begin{tabular}{llrrrr}
		\toprule
		 &  & Input & TC & TC-IS & Sp.up\tabularnewline
		\midrule 
		OSM & T {[}ms{]} & 258 & 14 & 9 & 27.7\tabularnewline
		~~-BaWü & P {[}$\cdot10^{3}${]} & 1504 & 80 & 48 & 31.5\tabularnewline
		 \addlinespace
		OSM & T {[}ms{]} & 2997 & 121 & 86 & 34.8\tabularnewline
		~~-Ger & P {[}$\cdot10^{3}${]} & 10229 & 595 & 354 & 28.9\tabularnewline
		 \addlinespace
		OSM & T {[}ms{]} & 32088 & 781 & 558 & 57.5\tabularnewline
		~~-Eur & P {[}$\cdot10^{3}${]} & 77933 & 3207 & 1928 & 40.4\tabularnewline
		 \addlinespace
		DIMACS & T {[}ms{]} & 2024 & 408 & 279 & 7.3\tabularnewline
		~~-Eur & P {[}$\cdot10^{3}${]} & 8965 & 1906 & 1153 & 7.8\tabularnewline
		 \addlinespace
		DIMACS & T {[}ms{]} & 3260 & 512 & 386 & 8.5\tabularnewline
		~~-US & P {[}$\cdot10^{3}${]} & 11885 & 2323 & 1502 & 7.9\tabularnewline
		\bottomrule
		\end{tabular}
		\par\end{centering}
	}
	\end{center}
\caption{\label{tab:cost-comp}Query performance with varying number of cost components on OSM-Ger with TopoCore-IS.}
\begin{centering}
\begin{tabular}{lrrrr}
\toprule
\#\,Costs &  8 & 16 & 32 & 64 \\
\midrule
Pop {[}$\cdot10^{3}${]} & 357 &  354 &348 & 340\\
Time [ms] & 80 & 108 & 132& 198 \\
\bottomrule
\end{tabular}
\par\end{centering}
\end{table}

\paragraph{Data Source Dependent Speedups}

The experimental results presented in Table~\ref{tab:topo-core} show that speedups achieved by our technique are significantly higher on OSM-based graphs (by a factor of up to 51.8/6.2 = 8.4). 
This is due to the significantly higher number of degree-2 nodes in these graphs, \cf~Table~\ref{tab:size}.
One may wonder whether this is a shortcoming of our technique. %

To the best of our knowledge, not many techniques have been evaluated on both OSM and non-OSM graphs, with the notable exception of~\cite{dgpw-crprn-13}, which has observed a similar effect: The speedup of their technique over Dijkstra's algorithm is up to 14.2 times higher on OSM than on non-OSM graphs.\footnote{They report speedups of 6\,093\,ms/1.67\,ms = 3\,649 on DIMACS-Eur, 6\,124\,ms/1.61\,ms = 3\,804 on DIMACS-US, 17\,750\,ms/1.98\,ms = 8\,965 on Bing data, but 77\,121\,ms/1.49\,ms = 51\,759 on their largest OSM graph. (Considering a route planning scenario different from ours.)}

These and our results suggest that OSM-based graphs are in some sense easier for speedup techniques compared to graphs with the same number of nodes but from other data sources. This needs to be considered in the comparison of different route planning techniques experimentally evaluated on road networks of different origin.

\paragraph{Additional Cost Components}

So far we have experimented with 8 cost components of 32 bits each.
However, some applications might require longer cost vectors.
We therefore perform additional query experiments on OSM-Germany with TopoCore-IS.
For these, we pad the existing cost vector with 8 components to 16, 32, and 64 components of 32 bits by adding random costs.
Table~\ref{tab:cost-comp} reports the average number of queue pop operations and running time.
The former is almost unaffected by the number of cost components.
However, the running time increases as more memory needs to be accessed.
Still, our approach scales very well: Going from 8 to 64 components requires 8 times more memory, but causes only a factor~2.5 increase in running time.

\subsection{Comparison with Related Work}
\begin{table*}
\caption{\label{tab:related} Comparison to related work. We report the number of criteria (\#\,Crit.) considered by each approach, the instance (in name and size) on which it was evaluated, the preprocessing time required, and the query time and speedup (over Dijkstra's algorithm) achieved. Where applicable we report customization time. We note if figures do not apply (---) or have not been reported (n/a). All timings are sequential, except for the GPU extension of CRP. CRP techniques were evaluated on an instance augmented with artificial U-turn costs. %
Differences in OSM graph size of the same instance are, to the best of our knowledge, due to different extraction dates.
}
\begin{centering}
\begin{tabular}{lrlrrrrrrr}
\toprule
 & & & |V|& |A| &Prepro.& Custom. & \multicolumn{2}{c}{Query}\\
Algorithm & \#\,Crit.  & Instance & [$\cdot 10^6$]& [$\cdot 10^6$]& [h:m:s] & [ms] & [ms] & Speedup\\
\midrule
CH~\cite{gssv-erlrn-12} & 1  & DIMACS-Eur &  18.0 & 42.2 & 2:45 & --- & 0.152 & n/a\\
CH, edge restrictions~\cite{DBLP:journals/jea/GeisbergerRST12} & 1  & NAVTEQ-US/CA  & 21.1 & 52.5 & 7:21:00 &--- & 1.18\phantom{0} & 2\,935\phantom{.0}\\
Pareto-SHARC~\cite{dw-pps-09} & 2  & DIMACS-Eur &  18.0 & 42.2 & 7:12:00 & --- & 35.4\phantom{00} & n/a \\
FlexCH~\cite{gks-rpfof-10} & 2  & DIMACS-Eur & 18.0 & 42.2 & 5:12:00 & --- & 0.98\phantom{0} & 6\,183\phantom{.0}\\
MultiCH~\cite{fs-pcchm-13} & 2  & OSM-BaWü** & 2.5 & 5.0 & 2:01 & ---  & 0.42\phantom{0} & 965\phantom{.0}\\
MultiCH~\cite{fs-pcchm-13} & 3  & OSM-BaWü** & 2.5 & 5.0 & 1:08 & --- & 3.16\phantom{0} & 234\phantom{.0}\\

\addlinespace
CRP~\cite{dgpw-crprn-13} & --- & DIMACS-Eur (Turn) &  18.0 & 42.2& 11:53 & 3\,770\phantom{.0} & 1.67\phantom{0} & 3\,649\phantom{.0}\\
CCH~\cite{dsw-cch-sea-14,hs-gbpo-15} & --- & DIMACS-Eur &  18.0 & 42.2& 4:40:41 & 2\,322\phantom{.5} & 0.27\phantom{0} & n/a \\
CRP on GPU~\cite{dkw-cddgp-14} & --- & DIMACS-Eur (Turn) &  18.0 & 42.2& 28:56  & 129.3 & 1.17\phantom{0} & n/a\\
\addlinespace
k-Path Cover~\cite{fns-opca-14} & 8  & OSM-BaWü* & 2.2 & 4.6 & 12\hide{12.4} & --- & 35\phantom{.000} & 10.8\\
k-Path Cover~\cite{fns-opca-14} & 8  & OSM-Ger* & 17.7 & 36.1 & 2:29\hide{149.4} & --- & 249\phantom{.000} & 13.1\\ 
TopoCore-IS & 8  & OSM-BaWü & 3.1 & 6.2 & 3\hide{2.7} & --- & 9\phantom{.000} & 27.7\\
TopoCore-IS & 8  & OSM-Ger & 20.7 & 41.8 & 35\hide{35.3} & --- & 86\phantom{.000} & 34.8\\
TopoCore-IS & 8  & DIMACS-Eur &  18.0 & 42.2& 36\hide{35.8}& --- & 279\phantom{.000} & 7.3\\
TopoCore-IS & 8  & DIMACS-US & 23.9 & 57.7 &43\hide{42.7} & --- & 386\phantom{.000} & 8.5\\
\bottomrule
\end{tabular}
\par\end{centering}
\end{table*}

While there is vast literature on route planning in road networks, most works consider query scenarios different from ours, making any direct comparison difficult.
We identify three classes of approaches related to the Personalized Route Planning~(PRP) scenario considered in our work: (1) adaptations of preprocessing techniques originally designed for fixed scalar costs, such as extensions of Contraction Hierarchies~(CH)~\cite{gssv-erlrn-12} that support multiple criteria~\cite{fs-pcchm-13,gks-rpfof-10} and arc restrictions (\eg, ``avoid highways'', vehicle weight limits, etc.) \cite{DBLP:journals/jea/GeisbergerRST12}, %
 or such as Pareto-SHARC~\cite{dw-pps-09}; (2)  Customizable Route Planning approaches~\cite{dgpw-crprn-13,dkw-cddgp-14,dsw-cch-sea-14}; (3) previous Personalized Route Planning approaches~\cite{fns-opca-14}. We report a detailed comparison of these approaches in Table~\ref{tab:related}. %

While plain CH (single fixed criterion, \ie, travel time) yields query times more than three orders of magnitude faster than ours, performance quickly degrades when considering arc restrictions or multiple criteria: While exact comparisons are difficult due to differences in benchmark instances, one roughly observes that considering arc restrictions as well as each additional criterion considered each decrease query speed by about an order of magnitude (0.152\,ms $\rightarrow$ 1.18\,ms, 0.152\,ms $\rightarrow$ 0.98\,ms, 0.42\,ms $\rightarrow$ 3.16\,ms). For three (somewhat correlated) criteria (distance, travel time, and fuel costs), CH performance on OSM-BaWü is already only factor 3--9 faster than for our approach in terms of query times and reported speedup~\cite{fs-pcchm-13}. 
This degradation of performance for more than two criteria likely means that the Contraction Hierarchies approach does not extend well to the PRP scenario considered in this work (an assessment also made by~\cite{fns-opca-14}). A similar, even stronger argument can be made against extending Pareto-SHARC~\cite{dw-pps-09} for PRP.

Customizable Route Planning~(CRP), introduced by~\cite{dgpw-crprn-13}, is closely related to PRP. However, instead of considering user preferences and restrictions as an input to each query, the cost of each arc (in the input graph as well as shortcuts) is established in a relatively quick \emph{customization} phase. In this phase, combinations of different criteria as well as restrictions (or live traffic delays) may be considered, but then, each subsequent query works on a single-criterion fixed metric. The original publication on CRP uses multi-level overlays and shortcuts~\cite{dgpw-crprn-13}, whereas CCH~\cite{dsw-cch-sea-14} is an adaption of CH to the customization setting. In~\cite{hs-gbpo-15} a better contraction order computation strategy is introduced resulting in the numbers of Table~\ref{tab:related}.  Directly applying both these techniques to PRP (by paying customization time for every change in user preferences), we observe that our approach to PRP outperforms them both, if user preferences change with every or up to every 8th query. (For perspective, recall the example of a fast route in the morning and a safe and fuel-efficient in the evening.)  
While customization can be parallelized on multiple CPU cores~\cite{dgpw-crprn-13,dsw-cch-sea-14}, only if it is highly parallelized on an external GPU~\cite{dkw-cddgp-14}, it becomes faster than our sequential queries. 
While having a GPU (for every concurrent user) is a strong assumption on the given computer hardware, we note that, even then, we achieve queries within the same order of magnitude (279\,ms compared to $129.3+1.17=130.47$\,ms).
Furthermore, in a server-setting, PRP-based approaches have no per-user memory consumption overhead (other than storing the objective function, if at all), whereas the per-user overhead for CRP and CCH depends on the graph size.

Finally, for a direct comparison for the Personalized Route Planning scenario, we contrast our results with those obtained by the k-Path Cover approach of~\cite{fns-opca-14} (which introduced the PRP scenario). 
On OSM graphs our PRP query speedup of 27.7.-57.5 more than doubles the maximum speedup of 13.2 previously achieved by~\cite{fns-opca-14}, while having lower preprocessing overhead. This observation is also supported by differences in absolute query runtime, even more so when considering the respective increase in OSM dataset size. Unfortunately, for their query experiments the authors of~\cite{fns-opca-14} focus exclusively on OSM graphs, hence we cannot compare on DIMACS graphs without speculation.

\section{Conclusions}%
We evaluated a preprocessing-based speedup technique for faster Personalized Route Planning. 
On all tested instances - which include very large-scale networks with hundreds of millions of nodes - we were able to achieve running times well below a second.
This is fast enough for many applications, including web services of moderate user base.
The main advantage of the Personalized Route Planning is that costs are individually adjusted for every user and every query in a very flexible way.
Rerunning preprocessing is only necessary when roads are build or cost vectors are adjusted (\eg, a new speed limit is posted).
We evaluated our technique both on OpenStreetMap data and on datasets from the 9th DIMACS implementation challenge, showing that it performs well on a large range of instances.

\balance

\end{document}